\documentclass[pra, twocolumn, showpacs, superscriptaddress,longbibliography]{revtex4-2}

\usepackage[pdftex,
            pdftitle={Deterministic preparation of squeezed Schrodinger cat states with arbitrary size and squeezing},
            pdfauthor={Oussama Houhou},
            pdfpagemode=UseOutlines,
            pdfstartview=FitH]{hyperref}
\usepackage{amsmath}
\usepackage{upgreek}
\usepackage{natbib}
\usepackage[utf8]{inputenc}
\usepackage[sans]{dsfont}
\usepackage{graphicx}
\usepackage{braket}
\usepackage{siunitx}

\newcommand{\diff}{\ensuremath{\mathrm{d}}}
\newcommand{\Hermite}{\ensuremath{\mathrm{H}}}
\newcommand{\hypergeom}{\ensuremath{{_1\mathrm{F}_1}}}
\newcommand{\hypergeomTwo}{\ensuremath{{_2\mathrm{F}_2}}}
\newcommand{\lrfloor}[1]{\ensuremath{{\left\lfloor #1\right\rfloor}}}

\newcommand{\dB}{\ensuremath{\mathrm{dB}}}

\newcommand{\refeq}[1]{Eq.~(\ref{#1})}
\newcommand{\Refeq}[1]{(\ref{#1})}
\newcommand{\refsec}[1]{Sec.~\ref{#1}}
\newcommand{\reffig}[1]{Fig.~\ref{#1}}

\newcommand{\EXP}[1]{{\mbox{e}}^{#1}}

\begin{document}

%-------------
%Front matter
%-------------

    \title{Deterministic preparation of squeezed Schr\"odinger cat states with arbitrary size and squeezing}
	\date{15 Novembre 2026}

	\author{Oussama Houhou}
	\email[E-mail: ]{houhou.oussama@univ-medea.dz}
	\affiliation{Laboratory of Physics of Experimental Techniques and Applications, University of Medea, Medea 26000, Algeria}

	\author{Hamza Adnane}
	\email[E-mail: ]{h.adnane@univ-bouira.dz}
	\affiliation{Physics Department, University of Bouira, Bouira 10000, Algeria}
    
    \begin{abstract}
		We introduce a scheme for the deterministic preparation of squeezed Schr\"odinger cat states based on reservoir engineering. Our proposal relies on the linear and quadratic coupling of an oscillator mode to an auxiliary mode that decays into a vacuum reservoir. For suitable couplings, we show that the system unconditionally relaxes into a pure steady state. We derive an analytical expression for the steady state in terms of the coupling constants, and show that it corresponds to a displaced and squeezed finite superposition of Fock states. Numerical simulations further confirms that this state is reached with unit fidelity from a thermal initial state, without measurement, feedback or post-selection. Additionally, we derive analytical expressions for the non-Gaussianity and the Wigner function of the steady state. Our results reveal that states exhibiting maximal non-Gaussianity are excellent approximations of squeezed Schr\"odinger cat states, with fidelities exceeding $0.996$. The optimal squeezing stays close to $3~\dB$, while the squared cat size scales proportionally with the highest Fock state occupation number in the finite superposition of the steady state. These maximally non-Gaussian states also maximise the Wigner logarithmic negativity. Finally, by selecting the appropriate ratios of the couplings, our scheme can generate cat states subject to arbitrary squeezing and displacement.
	\end{abstract}

    \maketitle
%-------------
%main matter
%-------------

%====================================================
%====================================================
%====================================================
\section{Introduction}\label{sec:introduction}
	Schr\"odinger cat states---quantum superpositions of macroscopically distinguishable states~\cite{schrodinger1935gegenwartige, brune1992manipulation, schleich1991nonclassical, yurke1986generating, brune1996observing, monroe1996schrodinger, ourjoumtsev2006generating, ourjoumtsev2007generation, vlastakis2013deterministically, leghtas2015confining, wang2016schrodinger, sanders2012review, gilchrist2004schrodinger, buvzek1992superpositions}---are among the most prominent nonclassical states in quantum mechanics, with applications spanning several areas of quantum science and technology. In continuous variables (CVs) quantum systems, a cat state is typically realized as a superposition of two coherent states with opposite amplitudes. Its size is characterized by the distance between the two components in phase space. The quantum interference arising from this superposition gives rise to oscillatory interference fringes in the corresponding Wigner function~\cite{wigner1932quantum} which manifest as negative regions in phase space~\cite{walschaers2021non, albarelli2018resource, kirchmair2013observation, ferrie2011quasi, deleglise2008reconstruction, ourjoumtsev2007generation}. Beyond their fundamental significance as paradigmatic manifestations of quantum coherence at the boundary between quantum and classical physics, cat states constitute valuable resources for a broad range of quantum technologies, including quantum computation~\cite{erkilicc2026unified, chamberland2022building, Eaton2022measurementbased, mirrahimi2014dynamically, leghtas2013hardware, lund2008fault, gilchrist2004schrodinger, ralph2003quantum, jeong2002efficient, cochrane1999macroscopically}, quantum communication and teleportation~\cite{neergaard2013quantum, jeong2001quantum}, and quantum metrology where they can enhance the precision of measurements~\cite{zheng2025quantum, bild2023schrodinger, tatsuta2019quantum, kwon2019nonclassicality, joo2011quantum, munro2002weak}.

	The role of cat states is particularly prominent in quantum computation with CVs~\cite{braunstein2005quantum}, where non-Gaussian resources, in conjunction with Gaussian states and operations, are essential for achieving universal quantum computation. For instance, a universal computation protocol introduced in Ref.~\cite{houhou2022unconditional} can be used to achieve universality through Gaussian measurements on entangled non-Gaussian states. Interestingly, cat states are both non-Gaussian and Wigner-negative, two properties that have been formally established as quantum resources~\cite{albarelli2018resource}. Moreover, large-amplitude squeezed cat states are robust resources for quantum error correction~\cite{erkilicc2026unified, winnel2024deterministic, chamberland2022building, Eaton2022measurementbased, leghtas2015confining, mirrahimi2014dynamically, leghtas2013hardware, lund2008fault, gottesman2001encoding, cochrane1999macroscopically}. Interfering multiple cat states also provides a route to the preparation of Gottesman--Kitaev--Preskill (GKP) states~\cite{winnel2024deterministic, gottesman2001encoding}. Additionally, squeezing offers advantages in the presence of excitation loss, allowing squeezed cat states to exhibit longer lifetimes than their unsqueezed counterparts owing to their lower average excitation number~\cite{takase2021generation}.

	Despite their considerable importance in quantum science and technology, the generation of cat states with sufficiently large amplitudes and high fidelities remains challenging, primarily due to their sensitivity to decoherence. Most established protocols, mainly in optics, are probabilistic---the target state is heralded by the outcome of a non-Gaussian measurement. Representative approaches include combining photon-number states with homodyne detection, which enabled the experimental generation of optical squeezed cat states~\cite{ourjoumtsev2007generation}; performing photon-number measurements on one mode of a Gaussian state, as a generalisation of photon subtraction from squeezed vacuum~\cite{takase2021generation}; and subtracting photons through networks of beam splitters and photon-number-resolving detectors. Although the latter approach can, in principle, produce cat states with large amplitudes and high fidelities, these improvements generally come at the expense of a reduced preparation success probability, which decreases sharply as the target fidelity increases~\cite{podoshvedov2023algorithm}.

	Deterministic schemes, which do not rely on post-selection, are relatively more recent and remain comparatively less explored. In optics, large-amplitude squeezed cat states can be prepared deterministically using Gaussian operations and photon-number measurements~\cite{winnel2024deterministic}, or through an all-optical dissipative scheme in which an engineered effective degenerate three-wave mixing induces the simultaneous loss of two photons~\cite{winnel2024deterministic, zhang2022all}. Beyond optical systems, cat-state generation has also been proposed in a variety of physical platforms, including optomechanical and other mechanical systems~\cite{naseem2026reservoir, hauer2023nonlinear, bild2023schrodinger, tan2013generation},  as well as microwave and superconducting systems~\cite{yang2025hot, li2024method, he2023fast, wang2022flying, zapletal2022stabilization, grimm2020stabilization, leghtas2015confining, mirrahimi2014dynamically,vlastakis2013deterministically}. Cat states can also be generated and preserved within CV cluster states by measurement-based protocols that supplement Gaussian processing with photon-number-resolving measurements~\cite{korolev2026growth, nesterova2026generation, renault2025squeezing, Eaton2022measurementbased, eaton2022generating, su2019conversion}.

	In this paper we introduce a scheme for the deterministic preparation of squeezed Schr\"odinger cat states based on engineered dissipation. Specifically, the target state is stabilised as the unique steady state of an open quantum system allowing it to be reached unconditionally, irrespective of the initial state and without measurements, feedback, or post-selection~\cite{diehl2008quantum,kraus2008preparation}. Our work builds on reservoir-engineering protocols for the deterministic preparation of Gaussian and non-Gaussian states, including entangled states~\cite{mcconnell2024unconditional,houhou2022unconditional,brunelli2018linear,brunelli2018unconditional,houhou2015generation}. We consider a harmonic oscillator whose dynamics is governed by a dissipative jump operator that is a combination of linear and quadratic terms in the ladder operators of the oscillator. Dissipators of this kind can be engineered, for instance, in cavity optomechanics, where an auxiliary damped cavity mode is coupled to the oscillator both linearly and quadratically in its displacement and is driven at multiple frequencies~\cite{houhou2022unconditional,brunelli2018linear,brunelli2018unconditional}. For suitable values of the couplings, the dynamics has a unique and pure steady state, which is a displaced and squeezed finite superposition of Fock states belonging to the family studied in Refs.~\cite{brunelli2018linear,brunelli2018unconditional}.

	These steady states are, up to a displacement, excellent approximations of squeezed Schr\"odinger cat states. The size of the cat and its squeezing are both tuned through the system parameters. We quantify the similarity with ideal squeezed cat states through the fidelity, which, upon optimisation over the size and squeezing of the target cat, exceeds $99\%$ for moderate cat sizes and approaches unity as the size increases. We also study the non-Gaussianity~\cite{genoni2008quantifying,adnane2019quantum} and the Wigner negativity~\cite{albarelli2018resource} of the steady states, and show that the cat states are those obtained when the non-Gaussianity is maximised. Finally, since Gaussian operations leave the Wigner negativity unchanged~\cite{albarelli2018resource} and protocols for performing arbitrary Gaussian operations on the prepared states are available~\cite{moore2017arbitrary}, the generated cat states can be further displaced, squeezed or processed without losing their non-Gaussian character.

	The paper is organised as follows. In \refsec{sec:system} we introduce the physical system, consisting of two bosonic modes coupled linearly and quadratically with one mode additionally coupled to a vacuum reservoir. The system's dynamics is described by a master equation. In \refsec{sec:steady-state} we investigate the existence and uniqueness of the steady state of the system, where we derive the conditions on the couplings for the system to evolve unconditionally towards this steady state. In \refsec{sec:displaced-squeezed-superposition} we find an analytical expression for the steady state and show that it is a finite superposition of Fock states, modulo displacement and squeezing operations. This steady state is parametrised by the highest number state ($n$) in the superposition, the squeezing strength ($s$), and the displacement ($x_0$). More Specifically, we will show that the finite superposition part of the  steady state of the system solely depends on $n$ and the ratio $\frac{x_0}{s}$. In \refsec{sec:system-dynamics} we numerically show that the dynamics reaches the aforementioned steady state. We show that the steady state is reached unconditionally with unit fidelity, starting from a thermal initial state. In \refsec{sec:Non-Gaussianity-measure} we obtain an analytical expression for the non-Gaussianity of the steady state and identify the states that maximise this quantity. In \refsec{sec:cat-steady-state} we show that these maximally non-Gaussian states closely approximate squeezed Schr\"odinger cat states, with fidelities exceeding $0.996$. Moreover, we show that the cat size scales proportionally to $\sqrt{n}$. In \refsec{sec:wigner-function-and-negativity} we derive an analytical expression for the Wigner function of the steady state and study its negativity through the Wigner logarithmic negativity. Finally, we summarise our findings and discuss possible directions for future research in \refsec{sec:conclusion}.

%====================================================
%====================================================
%====================================================
\section{The system}\label{sec:system}
\begin{figure}[hbt]
		\begin{center}
			\includegraphics[width=\linewidth]{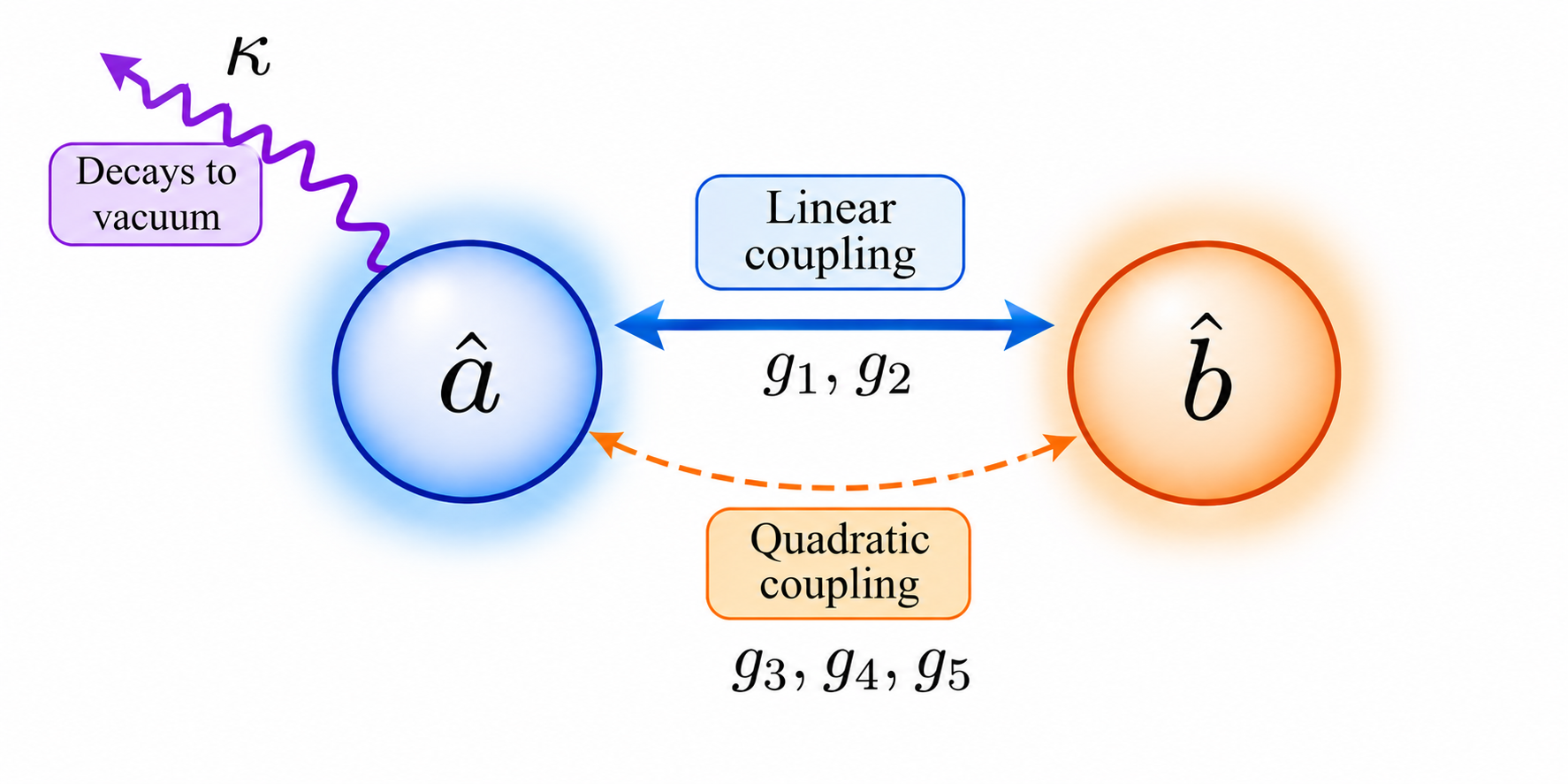}
			\caption{A system of two coupled bosonic modes---linearly and quadratically--and decaying to a cold reservoir.}\label{fig:the-system}
		\end{center}
	\end{figure}

	We consider the system depicted in \reffig{fig:the-system}. This system consists of two bosonic modes---characterised by their annihilation operators $\hat{a}$ and $\hat{b}$---interacting according to the Hamiltonian
	\begin{equation}
		\hat{H} = \hbar g \left(\hat{a}^\dagger\hat{f}\ +\ \hat{a}\hat{f}^\dagger\right)\ ,
	\end{equation}
	where the operator $\hat{f}$ is defined in terms of the oscillator's annihilation operator $\hat{b}$ by the following expression:
	\begin{equation}\label{eqn:operator-f-b}
		\hat{f}=g_1 \hat{b}+g_2 \hat{b}^\dagger+g_3 \hat{b}^2+g_4 \hat{b}^{\dagger 2}+g_5\left(\hat{b}\hat{b}^\dagger+\hat{b}^\dagger\hat{b}\right)\ .
	\end{equation}
	$g$ is the interaction rate between modes $\hat{a}$ and $\hat{f}$, and the coefficients $g_1$, $g_2$, $g_3$, $g_4$ and $g_5$ are complex dimensionless parameters which characterise the linear and quadratic interactions between the two bosonic modes. A detailed derivation of this Hamiltonian may be found in Ref.~\cite{houhou2022unconditional}. In this paper, we will consider $\hbar$ as the unit, i.e., $\hbar=1$.

	Additionally, the system is in interaction with a cold reservoir through mode $\hat{a}$. The dynamics of the system is described by a master equation of Lindblad form~\cite{lindblad1976generators, gorini1976completely, alicki2007quantum, gardiner2004quantum, breuer2002theory, carmichael1993open} as given below:
	\begin{equation}\label{eqn:master-equation-a-b}
		\dot{\hat\rho}_{ab}(t)=-i[\hat{H}, \hat{\rho}_{ab}(t)]\ +\ \kappa_a D[\hat{a}]\hat{\rho}_{ab}(t)\ ,
	\end{equation}
	where $\hat{\rho}_{ab}(t)$ is the system's density operator at time $t$, $\kappa_a$ is the decay rate of mode $\hat{a}$ into the vacuum, and $D[\cdot]$ is the super-operator defined for operator $\hat{\mathcal{O}}$ and density operator $\hat{\rho}$ as
	\begin{equation}
		D[\hat{\mathcal{O}}]\hat{\rho} = \hat{\mathcal{O}}\hat{\rho}\hat{\mathcal{O}}^\dagger-\frac12\hat{\mathcal{O}}^\dagger\hat{\mathcal{O}}\hat{\rho}-\frac12\hat{\rho}\hat{\mathcal{O}}^\dagger\hat{\mathcal{O}}\ .
	\end{equation}

	In a regime where the inter-bosonic couplings---$g\,g_1$,\ldots,$g\,g_5$---are much smaller than the decay rate to the reservoir, i.e., ${g\, g_j\ll\kappa_a}$ ($j=1,\ldots,5$), the mode $\hat{a}$ can be adiabatically eliminated~\cite{reiter2012effective,gardiner2004quantum,brunelli2018unconditional,woolley2014two,wiseman1993quantum}. Although this procedure is not mandatory for our theoretical results in this paper, it greatly simplifies calculations and requires less computational resources for the simulations. The master equation for the subsystem involving the mode $\hat{b}$ alone is written as
	\begin{equation}\label{eqn:master-equation}
		\dot{\hat\rho}(t)=\kappa\left(\hat{f}\hat\rho(t) \hat{f}^\dagger-\frac12\hat{f}^\dagger \hat{f}\hat\rho(t)-\frac12\hat\rho(t)\hat{f}^\dagger\hat{f}\right)\ ,
	\end{equation}
	with $\hat{\rho}$ is the density operator for the bosonic mode $\hat{b}$, and ${\kappa=4g^2/\kappa_a}$ is the new effective decay rate. This is a pure dissipative dynamics of the system.

%====================================================
%====================================================
%====================================================
\section{Criteria for an unconditional pure steady state}\label{sec:steady-state}
	In this section, we discuss the conditions that the couplings $g_1$, $g_2$, $g_3$, $g_4$ and $g_5$ in \refeq{eqn:operator-f-b} must satisfy in order for the system---governed by the master equation~\Refeq{eqn:master-equation}---to evolve unconditionally towards a unique pure steady state.

	We want the system to reach a steady state, $\ket{\varphi}$. Therefore, the operator $\hat{f}$ must satisfy the dark state condition~\cite{baumgartner2008analysisI,baumgartner2008analysisII,diehl2008quantum,kraus2008preparation},
	\begin{equation}\label{eqn:dark-state-condition}
		\hat{f}\ket{\varphi}=0\ .
	\end{equation}
	In terms of position and momentum quadratures, $\hat{x}$ and $\hat{p}$ respectively, the operator $\hat{f}$ is given by
	\begin{equation}\label{eqn:operator-f-x-p}
		\hat{f}=-c_1\hat{p}^2+i(c_2 \hat{x}+c_3)\hat{p}+c_4 \hat{x}^2+c_5 \hat{x}+\frac12 c_2\ ,
	\end{equation}
	with
	\begin{subequations}\label{eqn:c1--c5}
		\begin{align}
			c_1	&=	-g_5+\frac12(g_3+g_4)\ ,\\
			c_2	&=	g_3-g_4\ ,\\
			c_3	&=	\frac{1}{\sqrt2}(g_1-g_2) \ ,\\
			c_4	&=	g_5+\frac12(g_3+g_4)\ ,\\
			c_5	&=	\frac{1}{\sqrt2}(g_1+g_2)\ .
		\end{align}
	\end{subequations}
	In position representation, $\hat{x}\equiv x$ and $\hat{p}\equiv -i\frac{\diff}{\diff x}$, condition~\Refeq{eqn:dark-state-condition} becomes
	\begin{equation}\label{eqn:diff-eq-wave-eq-general}
		c_1\varphi''+(c_2 x+c_3)\varphi'-\left(c_4x^2+c_5x+\frac12 c_2\right)\varphi=0\ .
	\end{equation}
	Furthermore, we want the system to reach a unique pure steady state. A sufficient condition to fulfil this requirement is to set $c_1$ to zero,
	\begin{equation}\label{eqn:c1=0-condition}
		c_1 = 0\ .
	\end{equation}
	It is worth mentioning that there are other sets of parameters that also lead to a unique pure steady state, e.g., see Refs.~\cite{houhou2022unconditional,brunelli2018linear,brunelli2018unconditional}. Under the constraint \Refeq{eqn:c1=0-condition}, we want to find the solution of the following differential equation,
	\begin{equation}\label{eqn:diff-eq-wave-eq-first-order}
		(c_2 x+c_3)\varphi'-\left(c_4x^2+c_5x+\frac12 c_2\right)\varphi=0\ ,
	\end{equation}
	where the two parameters $c_2$ and $c_3$ must not be simultaneously zero; otherwise, there will be no physical steady state.

	It is worth noting that not all values of $c_2$, $c_3$, $c_4$, and $c_5$ will give a physical state, i.e. with square-integrable function $\varphi$. In \refsec{sec:displaced-squeezed-superposition}, we give particular values for these parameters that ensure the physicality of the state.

%====================================================
%====================================================
%====================================================
\section{Displaced Squeezed Finite Superposition of Fock States}\label{sec:displaced-squeezed-superposition}
	We find in the following the solutions of the differential equation~\Refeq{eqn:diff-eq-wave-eq-first-order}. We choose the following configuration
	\begin{subequations}\label{eqn:c3--c5}
		\begin{align}
			c_3	&=	\alpha c_2\ ,\\
			c_4	&=	\frac{1}{s^2} c_2\ ,\\
			c_5	&=	\beta c_2\ ,
		\end{align}
	\end{subequations}
	where $\alpha$, $\beta$, and $s$ are real parameters with $c_2$ and $s$ are non-zero. The solution of the differential equation~\Refeq{eqn:diff-eq-wave-eq-first-order} takes the form
	\begin{equation}
		\varphi(x)=N_\varphi(x+\alpha)^{-\left(\frac{\alpha^2}{s^2}-\alpha\beta+\frac12\right)}\EXP{\frac{-1}{2s^2}\left(x+\beta s^2-\alpha\right)^2}\ ,
	\end{equation}
	where $N_\varphi$ is an integration constant.

	A sufficient condition for the physicality of the steady state $\varphi$ is that the exponent, $\left(-\frac{\alpha^2}{s^2}+\alpha\beta-\frac12\right)$, is a positive integer. We define $n$ and $x_0$ by
	\begin{align}
		n	&=	\alpha\beta-\frac{\alpha^2}{s^2}-\frac12\ ,\label{eqn:n-alpha-beta}\\
		x_0	&=	\alpha-\beta s^2\ .\label{eqn:x0-alpha-beta}
	\end{align}
	Note that $n$ in terms of $x_0$ reads $n=\frac{-\alpha}{s^2} x_0-\frac12$. It is easy to conclude that the requirement $n\in\mathds{N}$ ensures that both $x_0$ and $\alpha$ are non-zero. In terms of $n$, $s$, and $x_0$, the steady state wave function, $\varphi\equiv\varphi_{n,s,x_0}$, can be written as
	\begin{equation}\label{eqn:wave-function-phi}
		\varphi_{n,s,x_0}(x)=N_\varphi\left(x-\frac{s^2\left(n+\frac12\right)}{x_0}\right)^n\EXP{\frac{-1}{2s^2}(x-x_0)^2}\ .
	\end{equation}

	Notice that the form of the argument of the exponential function appearing in \refeq{eqn:wave-function-phi} suggests that the state with the wave function $\varphi_{n,s,x_0}(x)$ is in fact the result of the action of the displacement and squeezing operators on another state with wave function $\psi_{n,s,x_0}(x)$, as shown by the following expression:
	\begin{align}
		\varphi_{n,s,x_0}(x)
		&=\braket{x|\varphi_{n,s,x_0}}\nonumber\\
		&=\braket{x|\hat{D}\left(\frac{x_0}{\sqrt2}\right)\hat{S}(s)|\psi_{n,s,x_0}}\ ,\label{eqn:steady-state-phi-n-s-x0}
	\end{align}
	where $\hat{D}(\lambda)$ and $\hat{S}(s)$ are the displacement and squeezing operators respectively, defined for the complex number $\lambda$ (the displacement parameter) and the non-zero positive number $s$ (the squeezing parameter) by
	\begin{subequations}
		\begin{align}
			\hat{D}(\lambda)	&=\EXP{\lambda \hat{b}^\dagger-\lambda^* \hat{b}}\ ,\\
			\hat{S}(s)		&=\EXP{\frac12\ln s\ \left(\hat{b}^{\dagger 2}-\hat{b}^2\right)}\ .
		\end{align}
	\end{subequations}
	The wave function $\psi_{n,s,x_0}(x)\equiv\braket{x|\psi_{n,s,x_0}}$ reads
	\begin{equation}
		\psi_{n,s,x_0}(x)=N\left(x+\xi\right)^n\ \EXP{-\frac12 x^2}\ ,
	\end{equation}
	where $N$ is a normalisation factor, and
	\begin{equation}
		\xi	=	\frac{x_0}{s}-\frac{n+\frac12}{\frac{x_0}{s}}\ .
	\end{equation}
	Note that the wave function $\psi_{n,s,x_0}$ depends on both $n$ and the ratio $x_0/s$. Therefore, it is more convenient and significant to change the notation from $\ket{\psi_{n,s,x_0}}$ to ${\ket{\psi_{n,\eta}}\equiv\ket{n,\eta}}$, with
	\begin{equation}\label{eqn:x-0-s-eta}
		\eta=\frac{x_0}{s}\ .
	\end{equation}
	The wave function is now written as
	\begin{equation}
		\psi_{n,\eta}(x)=N_{n,\eta}(x+\xi_{n,\eta})^n\ \EXP{-\frac12 x^2}\ ,
	\end{equation}
	with $\xi_{n,\eta}$ is now defined as
	\begin{equation}
		\xi_{n,\eta}=\eta-\frac{n+\frac12}{\eta}\ .
	\end{equation}
	The normalisation constant $N_{n,\eta}$ is calculated from the condition $\braket{n,\eta|n,\eta}=1$,
	\begin{align}
		&\braket{n,\eta|n,\eta}=1\ \Leftrightarrow\ |N_{n,\eta}|^2\int\limits_{-\infty}^\infty\ (x+\xi_{n,\eta})^{2n}\EXP{-x^2}\ \diff x=1\ ,\nonumber\\[1em]
		&\Rightarrow
		|N_{n,\eta}|=\frac{1}{\sqrt{\Gamma\left(n+\frac12\right)\ \hypergeom\left(-n;\frac12;-\xi_{n,\eta}^2\right)}}\ ,
	\end{align}
	and the final form of the wave function reads (taking $N_{n,\eta}$ to be real)
	\begin{equation}\label{eqn:wave-function-psi}
		\psi_{n,\eta}(x)=\frac{(x+\xi_{n,\eta})^n\ \EXP{-\frac12 x^2}}{\sqrt{\Gamma\left(n+\frac12\right)\ \hypergeom\left(-n;\frac12;-\xi_{n,\eta}^2\right)}}\ .
	\end{equation}
	In this formula, $\Gamma$ is the gamma function and $\hypergeom$ is the Kummer confluent hypergeometric function of the first kind~\cite{olver2010nist}.

	Up to now, we showed that the steady state of the system having a dynamics described by the master equation~\Refeq{eqn:master-equation} is $\ket{n,\eta}$, up to displacement and squeezing operations, i.e.,
	\begin{equation}\label{eqn:ket-varphi_n_s_x0}
		\ket{\varphi_{n, s, x_0}}=\hat{D}\left(\frac{x_0}{\sqrt2}\right)\hat{S}(s)\ket{n,\eta}\ .
	\end{equation}
	In the following we proceed and show that this state can be written as a finite superposition of Fock states.

	The wave function $\psi_{n,\eta}$ written in \refeq{eqn:wave-function-psi} can be put in the following form,
	\begin{equation}
		\psi_{n,\eta}(x)=N_{n,\eta}\ \sum\limits_{k=0}^n\ \frac{n!\ \xi_{n,\eta}^{n-k}}{k!(n-k)!}\ x^k\EXP{\frac{-1}{2}x^2}\ .
	\end{equation}
	Taking into account that the wave function of the Fock state $\ket{\ell}$ is
	\begin{equation}
		\phi_\ell(x)=\frac{1}{\uppi^\frac14\sqrt{2^\ell\ell!}}\ \Hermite_\ell(x)\ \EXP{\frac{-1}{2}x^2}\ ,
	\end{equation}
	where $\Hermite_\ell$ is the $\ell^\text{th}$-order Hermite polynomial, and the identity~\cite{abramowitz1972handbook}
	\begin{equation}
		x^k=\frac{k!}{2^k}\ \sum\limits_{m=0}^\lrfloor{\frac{k}{2}}\ \frac{1}{m!(k-2m)!}\ \Hermite_{k-2m}(x)\ ,
	\end{equation}
	we arrive at the following expression for the steady state wave function,
	\begin{multline}
		\psi_{n,\eta}(x)= n! \uppi^\frac14 N_{n,\eta}\\
		\times\ \sum\limits_{k=0}^n\sum\limits_{m=0}^\lrfloor{\frac{k}{2}}\ \frac{\xi_{n,\eta}^{n-k}\ \phi_{k-2m}(x)}{2^m m! (n-k)! \sqrt{2^k(k-2m)!}}\ .
	\end{multline}
	Equivalently, the corresponding ket vector for the steady state can be written as
	\begin{multline}
		\ket{n,\eta}=n! \uppi^\frac14 N_{n,\eta}\\
		\times\sum\limits_{k=0}^n\sum\limits_{m=0}^\lrfloor{\frac{k}{2}}\frac{\xi_{n,\eta}^{n-k}}{2^m m! (n-k)! \sqrt{2^k(k-2m)!}}\ket{k-2m}\ ,
	\end{multline}
	where we used the notation $\phi_\ell(x)\equiv\braket{x|\ell}$. The expression of $\ket{n,\eta}$ is simplified further to
	\begin{multline}\label{eqn:ket-n-eta}
		\ket{n,\eta}=\frac{n!}{\sqrt{2^n (2n-1)!!\ \hypergeom\left(-n;\frac12;-\xi_{n,\eta}^2\right)}}\\
		\times\Big(\ket{n,\eta}_\text{even}+\ket{n,\eta}_\text{odd}\Big)\ ,
	\end{multline}
	where the even and odd states $\ket{n,\eta}_\text{even}$ and $\ket{n,\eta}_\text{odd}$ write
	\begin{subequations}
		\begin{align}
			\ket{n,\eta}_\text{even} &= \sum\limits_{k=0}^\lrfloor{\frac{n}{2}}\frac{2^k\left(2\xi_{n,\eta}\right)^{\upepsilon_n}}{\left(\lrfloor{\frac{n}{2}}-k\right)!\ \sqrt{(2k)!}}\nonumber\\
			&\times\hypergeom\left(k-\lrfloor{\frac{n}{2}};\frac12+\upepsilon_n;-\xi_{n,\eta}^2\right)\ket{2k}\ ,\\
			\ket{n,\eta}_\text{odd} &= \sum\limits_{k=0}^\lrfloor{\frac{n-1}{2}}\frac{2^{k+\frac12}\left(2\xi_{n,\eta}\right)^{1-\upepsilon_n}}{\left(\lrfloor{\frac{n-1}{2}}-k\right)!\ \sqrt{(2k+1)!}}\nonumber\\
			&\times\hypergeom\left(k-\lrfloor{\frac{n-1}{2}};\frac32-\upepsilon_n;-\xi_{n,\eta}^2\right)\ket{2k+1}\ .
		\end{align}
	\end{subequations}
	The symbol $\upepsilon_n$ denotes the remainder of the division of $n$ by $2$,
	\begin{equation}
		\upepsilon_n\equiv \mod(n,2)=\left\{
		\begin{aligned}
			&0	&&\text{if }n\text{ is even}\\
			&1	&&\text{if }n\text{ is odd}
		\end{aligned}
		\right.\ .
	\end{equation}

%====================================================
%====================================================
%====================================================
\section{The dynamics of the system}\label{sec:system-dynamics}
	In this section, we numerically confirm that the system unconditionally reaches the steady state given in \refeq{eqn:ket-varphi_n_s_x0}. First, we find the couplings $g_1$, \ldots, $g_5$ that ensure the dynamics of the system reaches the steady state~\Refeq{eqn:ket-varphi_n_s_x0}. Taking into account equations~\Refeq{eqn:c1--c5}, condition~\Refeq{eqn:c1=0-condition}, equations~\Refeq{eqn:c3--c5}, \refeq{eqn:n-alpha-beta}, and \refeq{eqn:x0-alpha-beta}, we arrive at the following,
	\begin{subequations}
		\begin{align}
			g_2	&=	g_1\ \frac{1-s^2+\frac{(\frac{x_0}{s})^2}{n+\frac12}}{1+s^2+\frac{(\frac{x_0}{s})^2}{n+\frac12}}\ ,\\
			g_3	&=	g_1\ \frac{\frac{-1}{\sqrt2}\left(s+\frac{1}{s}\right)+\frac{\frac{x_0}{s}}{n+\frac12}}{1+s^2+\frac{(\frac{x_0}{s})^2}{n+\frac12}}\ ,\\
			g_4	&=	g_1\ \frac{\frac{-1}{\sqrt2}\left(s-\frac{1}{s}\right)+\frac{\frac{x_0}{s}}{n+\frac12}}{1+s^2+\frac{(\frac{x_0}{s})^2}{n+\frac12}}\ ,\\
			g_5	&=	g_1\ \frac{\frac{-1}{\sqrt2\ s}\ \frac{\frac{x_0}{s}}{n+\frac12}}{1+s^2+\frac{(\frac{x_0}{s})^2}{n+\frac12}}\ .
		\end{align}
	\end{subequations}
	In terms of $\eta$, the couplings are given by
	\begin{subequations}
		\begin{align}
			g_2	&=	g_1\ \frac{1-s^2+\frac{\eta^2}{n+\frac12}}{1+s^2+\frac{\eta^2}{n+\frac12}}\ ,\\
			g_3	&=	g_1\ \frac{\frac{-1}{\sqrt2}\left(s+\frac{1}{s}\right)+\frac{\eta}{n+\frac12}}{1+s^2+\frac{\eta^2}{n+\frac12}}\ ,\\
			g_4	&=	g_1\ \frac{\frac{-1}{\sqrt2}\left(s-\frac{1}{s}\right)+\frac{\eta}{n+\frac12}}{1+s^2+\frac{\eta^2}{n+\frac12}}\ ,\\
			g_5	&=	g_1\ \frac{\frac{-1}{\sqrt2\ s}\ \frac{\eta}{n+\frac12}}{1+s^2+\frac{\eta^2}{n+\frac12}}\ .
		\end{align}
	\end{subequations}
	Next, we write the master equation~\Refeq{eqn:master-equation} in a dimensionless form suitable for numerical simulations. For that purpose, we rescale the field operator $\hat{f}$ by the factor $\sqrt{g_1^2-g_2^2}$,
	\begin{equation}
		\hat{f'} = \frac{1}{\sqrt{g_1^2-g_2^2}}\ \hat{f}\ .
	\end{equation}
	We may write the new operator $\hat{f'}$ as
	\begin{equation}
		\hat{f'}=g'_1 \hat{b}+g'_2 \hat{b}^\dagger+g'_3 \hat{b}^2+g'_4 \hat{b}^{\dagger 2}+g'_5\left(\hat{b}\hat{b}^\dagger+\hat{b}^\dagger\hat{b}\right)\ ,
	\end{equation}
	where the couplings $g'_1$, \ldots, $g'_5$ are given by
	\begin{subequations}
		\begin{align}
			g'_1	&=	\frac{1}{2s}\ \frac{1+s^2+\frac{\eta^2}{n+\frac12}}{\sqrt{1+\frac{\eta^2}{n+\frac12}}}\ ,\\
			g'_2	&=	\frac{1}{2s}\ \frac{1-s^2+\frac{\eta^2}{n+\frac12}}{\sqrt{1+\frac{\eta^2}{n+\frac12}}}\ ,\\
			g'_3	&=	\frac{-1}{2\sqrt2\ s^2}\ \frac{(1+s^2)\ \frac{\eta}{n+\frac12}}{\sqrt{1+\frac{\eta^2}{n+\frac12}}}\ ,\\
			g'_4	&=	\frac{-1}{2\sqrt2\ s^2}\ \frac{(1-s^2)\ \frac{\eta}{n+\frac12}}{\sqrt{1+\frac{\eta^2}{n+\frac12}}}\ ,\\
			g'_5	&=	\frac{-1}{2\sqrt2\ s^2}\ \frac{\frac{\eta}{n+\frac12}}{\sqrt{1+\frac{\eta^2}{n+\frac12}}}\ .
		\end{align}
	\end{subequations}
	Finally, the dimensionless master equation is given by
	\begin{equation}\label{eqn:master-equation-dimensionless-bis}
		\frac{\diff\hat{\rho}(t')}{\diff t'}=\left(\hat{f'}\hat{\rho}(t') \hat{f'}^\dagger-\frac12\hat{f'}^\dagger \hat{f'}\hat{\rho}(t')-\frac12\hat{\rho}(t')\hat{f'}^\dagger\hat{f'}\right)\ ,
	\end{equation}
	where the dimensionless parameter $t'=\frac{\kappa t}{g_1^2-g_2^2}$ is given by
	\begin{equation}
		t' = \frac{\kappa t}{4s^2 g_1^2}\ \frac{1+s^2+\frac{\eta^2}{n+\frac12}}{1+\frac{\eta^2}{n+\frac12}} \ .
	\end{equation}

	Alternatively, we can consider a different, but equivalent, master equation whose steady state is just the finite superposition of Fock states, i.e., $\ket{n, \eta}$ given in \refeq{eqn:ket-n-eta}. Since the steady state is a displaced and squeezed finite superposition of Fock states---as shown in \refeq{eqn:ket-varphi_n_s_x0}---we can find the master equation that has the finite superposition as the steady state. This master equation is given by
	\begin{equation}\label{eqn:master-equation-transformed-Displ-Sqz}
		\frac{\diff\hat{\mathcal{R}}(t')}{\diff t'}=\hat{\tilde{f}}\hat{\mathcal{R}}(t') \hat{\tilde{f}}^\dagger-\frac12\hat{\tilde{f}}^\dagger \hat{\tilde{f}}\hat{\mathcal{R}}(t')-\frac12\hat{\mathcal{R}}(t')\hat{\tilde{f}}^\dagger\hat{\tilde{f}}\ .
	\end{equation}
	$\hat{\mathcal{R}}$ and $\hat{\tilde{f}}$ are, respectively, the transformed state of the system and the field operator by the displacement $\hat{D}(x_0/\sqrt2)$ and the squeezing $\hat{S}(s)$ operators,
	\begin{subequations}
		\begin{align}
			\hat{\mathcal{R}} &= \hat{U}^\dagger\ \hat{\rho}\ \hat{U}\ ,\\
			\hat{\tilde{f}} &= \hat{U}^\dagger\ \hat{f'}\ \hat{U}\ ,
		\end{align}
	\end{subequations}
	where $\hat{U}$ is the aforementioned transformation,
	\begin{equation}
		\hat{U} = \hat{D}\left(\frac{x_0}{\sqrt2}\right)\ \hat{S}(s)\ .
	\end{equation}
	The new field operator, $\hat{\tilde{f}}$, reads
	\begin{equation}
		\hat{\tilde{f}} = \tilde{g}_1 \hat{b}+\tilde{g}_2 \hat{b}^\dagger+\tilde{g}_3 \hat{b}^2+\tilde{g}_4 \hat{b}^{\dagger 2}+\tilde{g}_5 \hat{b}^\dagger\hat{b}+\tilde{g}_6\mathds{1}\ .
	\end{equation}
	$\mathds{1}$ in this expression denotes the identity operator acting on the Hilbert space of the oscillator. The new couplings are given by
	\begin{subequations}
		\begin{align}
			\tilde{g}_1	&=	\frac{1-\frac{\eta^2}{n+\frac12}}{\sqrt{1+\frac{\eta^2}{n+\frac12}}}\ ,\\
			\tilde{g}_2	&=	0\ ,\\
			\tilde{g}_3	&=	\frac{-1}{\sqrt2}\ \frac{\frac{\eta}{n+\frac12}}{\sqrt{1+\frac{\eta^2}{n+\frac12}}}\ ,\\
			\tilde{g}_4	&=	0\ ,\\
			\tilde{g}_5	&=	\frac{-1}{\sqrt2}\ \frac{\frac{\eta}{n+\frac12}}{\sqrt{1+\frac{\eta^2}{n+\frac12}}}\ ,\\
			\tilde{g}_6	&=	\frac{1}{\sqrt2}\ \frac{n\ \frac{\eta}{n+\frac12}}{\sqrt{1+\frac{\eta^2}{n+\frac12}}}\ .
		\end{align}
	\end{subequations}

	The master equation \Refeq{eqn:master-equation-transformed-Displ-Sqz} can be put in the form
	\begin{equation}\label{eqn:master-equation-transformed}
		\frac{\diff\hat{\mathcal{R}}(\tau)}{\diff\tau}=\hat{F}\hat{\mathcal{R}}(\tau) \hat{F}^\dagger-\frac12\hat{F}^\dagger \hat{F}\hat{\mathcal{R}}(\tau)-\frac12\hat{\mathcal{R}}(\tau)\hat{F}^\dagger\hat{F}\ ,
	\end{equation}
	where we defined the operator $\hat{F}$ by
	\begin{equation}
		\hat{\tilde{f}} = \frac{-1}{\sqrt2\ \sqrt{1+\frac{\eta^2}{n+\frac12}}}\ \hat{F}\ ,
	\end{equation}
	with the expression
	\begin{equation}
		\hat{F}	= \Big[-\sqrt2\ \frac{n+\frac12}{\eta}\left(1-\frac{\eta^2}{n+\frac12}\right)\hat{b} +  \hat{b}^2+\hat{b}^\dagger \hat{b}-n\mathds{1}\Big]\ .
	\end{equation}
	Additionally, we defined the dimensionless parameter $\tau$ by
	\begin{equation}
		\tau = \frac{t'}{2}\ \frac{\frac{\eta^2}{(n+\frac12)^2}}{1+\frac{\eta^2}{n+\frac12}}\ .
	\end{equation}

	Now we numerically confirm that the system reaches the steady state $\ket{n, \eta}$. We proceed as follows: A target state is fixed by setting the parameters $n$, $s$, and $x_0$. Then, we numerically solve the master equation governing the system's dynamics, \refeq{eqn:master-equation-transformed}, record the quantum state at time $t$, and calculate the fidelity~\cite{uhlmann1976transition, jozsa1994fidelity} between the state of the system at time $t$ and the target steady state.
	
\begin{figure}[hbt]
		\begin{center}
			\includegraphics[width=\linewidth]{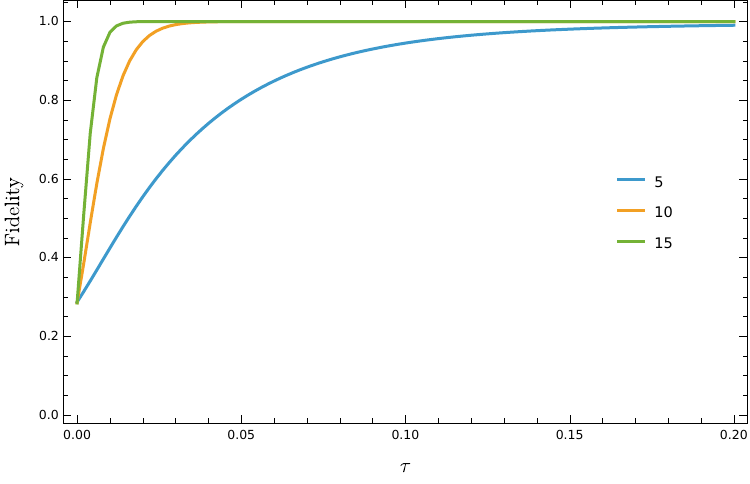}
			\caption{Fidelity as a function of evolution time for different values of $n$: $n=5,\ 10,\ 15$. The target steady state corresponds to $n$, $s=1$, and $x_0=1$.}\label{fig:fidel-time-n}
		\end{center}
	\end{figure}

	The results for $s=1$ (i.e. no squeezing), $x_0=1$, different values of $n$ ($5$, $10$, and $15$), and a thermal initial state are shown in \reffig{fig:fidel-time-n}. It is worth noting that we could choose any other values for $s$ and $x_0$ and the results remain the same. Those particular values of $s$ and $x_0$ had been chosen to make the computation more efficient in terms of time and memory resources. From the shown plots, we notice that the fidelity corresponding to all target states always reaches unity, which confirms our previously stated result, i.e. the system is unconditionally relaxed to the target state $\ket{n, \eta}$. Moreover, the rate of reaching the steady state varies with $n$ for constant $\eta$. This is justified by the fact that for increasing values of $n$, the couplings $g_j$ ($j=1,\ldots,5$) increase accordingly---see equation sets~\Refeq{eqn:c1--c5}, \Refeq{eqn:c3--c5}, and~\Refeq{eqn:n-alpha-beta}. Consequently, the system approaches its steady state more quickly.

%====================================================
%====================================================
%====================================================
\section{Non-Gaussianity measure}\label{sec:Non-Gaussianity-measure}
	In the following, we give an analytical expression for the non-Gaussianity measure of the state $\ket{n,\eta}$. Namely, we will be using the corresponding wave function $\psi_{n,\eta}$ given by \refeq{eqn:wave-function-psi}.

	%====================================================
	%====================================================
	%====================================================
	\subsection{Covariance matrix}\label{sec:Non-Gaussianity-measure-cov-mat}
		The elements of the covariance matrix $\sigma^{n,\eta}$ for the state $\ket{n,\eta}$ write~\cite{weedbrook2012gaussian, adesso2014continuous, ferraro2005gaussian}
		\begin{equation}
			\sigma_{ij} = \frac12\braket{\{\hat{R_i},\hat{R_j}\}} - \braket{\hat{R_i}}\braket{\hat{R_j}}\ ,
		\end{equation}
		where $\hat{R} = \left(\hat{x},\hat{p}\right)^\top$ is the vector of the quadrature operators.

		In order to write the expression of the covariance matrix $\sigma$, one needs to evaluate the expectation values for the quadrature operators, $\hat{x}$ and $\hat{p}$, and their squares, $\hat{x}^2$ and $\hat{p}^2$, in the state $\ket{n,\eta}$. For this, the average value of an operator $\hat{\mathcal{O}}$ is obtained as
		\begin{equation}
			\braket{\hat{\mathcal{O}}} = \braket{n,\eta|\hat{\mathcal{O}}|n,\eta} = \int\limits_{-\infty}^\infty\psi_{n,\eta}(x)\hat{\mathcal{O}}_x\psi_{n,\eta}(x)\ \diff x,
		\end{equation}
		where $\mathcal{O}_x$ denotes the $x$-representation of the operator $\hat{\mathcal{O}}$. We obtain the following expectation values:
		\begin{widetext}
			\begin{subequations}
				\begin{align}
					\braket{\hat{x}}		&=	\int\limits_{-\infty}^\infty\ x\Big(\psi_{n,\eta}(x)\Big)^2\ \diff x=	2n\xi_{n,\eta}\frac{\hypergeom(1-n;\frac32;-\xi_{n,\eta}^2)}{\hypergeom(-n;\frac12;-\xi_{n,\eta}^2)}\ ,\\
					\braket{\hat{x}^2}	&=	\int\limits_{-\infty}^\infty\ x^2\Big(\psi_{n,\eta}(x)\Big)^2\ \diff x=	\left(n+\frac12\right)\frac{\hypergeomTwo(\frac12-n,-n;\frac12,\frac{-1}{2}-n;-\xi_{n,\eta}^2)}{\hypergeom(-n;\frac12;-\xi_{n,\eta}^2)}\ ,\\
					\braket{\hat{p}}		&=	-i\int\limits_{-\infty}^\infty\ \psi_{n,\eta}(x)\psi'_{n,\eta}(x)\ \diff x
					=	0\ ,\\
					\braket{\hat{p}^2}	&=	-\int\limits_{-\infty}^\infty\ \psi_{n,\eta}(x)\psi''_{n,\eta}(x)\ \diff x\nonumber\\
					&=	\left(\frac12 + \frac{1}{1+2\xi_{n,\eta}^2}\right)\delta_{n,1} + (1-\delta_{n,1})\left[1+\frac{1+2\xi_{n,\eta}^2}{2(2n-1)}-\frac{2n+1}{2n-1}\xi_{n,\eta}^2\frac{\hypergeom(-n;\frac32;-\xi_{n,\eta}^2)}{\hypergeom(-n;\frac12;-\xi_{n,\eta}^2)}\right]\\
					\braket{\hat{x}\hat{p}}		&=	-i\int\limits_{-\infty}^\infty\ x\psi_{n,\eta}(x)\psi'_{n,\eta}(x)\ \diff x\nonumber\\
					&\Rightarrow\braket{\hat{x}\hat{p}+\hat{p}\hat{x}} = 0\ .
				\end{align}
			\end{subequations}
		\end{widetext}

		We obtained zero expectation values for both $\braket{\hat{p}}$ and $\braket{\hat{x}\hat{p}}$. Consequently, the off-diagonal elements of the covariance matrix for the state $\ket{n,\eta}$ are zero for all values of $n$ and $\eta$. The covariance matrix takes the diagonal form
		\begin{equation}\label{eqn:cov-mat}
			\sigma^{n,\eta} =
			\begin{pmatrix}
				\sigma_{11}	&	0\\
				0			&	\sigma_{22}
			\end{pmatrix}
			\ ,
		\end{equation}
		where $\sigma_{11}=\braket{\hat{x}^2}-\braket{\hat{x}}^2$ and $\sigma_{22}=\braket{\hat{p}^2}-\braket{\hat{p}}^2$ are calculated and expressed by
		\begin{widetext}
		\begin{subequations}\label{eqn:cov-mat-11-22}
			\begin{align}
				\sigma_{11}	&=	n+\frac12-\frac{4n(n+\frac12)\xi_{n,\eta}^2\,\hypergeom(1-n;\frac32;-\xi_{n,\eta}^2)\,\hypergeom(-n;\frac32;-\xi_{n,\eta}^2)}{\Big(\hypergeom(-n;\frac12;-\xi_{n,\eta}^2)\Big)^2}\ ,\\
				\sigma_{22}	&=	\left(\frac12 + \frac{1}{1+2\xi_{n,\eta}^2}\right)\delta_{n,1}+ (1-\delta_{n,1})\left[1+\frac{1+2\xi_{n,\eta}^2}{2(2n-1)}-\frac{2n+1}{2n-1}\xi_{n,\eta}^2\frac{\hypergeom(-n;\frac32;-\xi_{n,\eta}^2)}{\hypergeom(-n;\frac12;-\xi_{n,\eta}^2)}\right]\ .
			\end{align}
		\end{subequations}
	\end{widetext}

	%====================================================
	%====================================================
	%====================================================
	\subsection{Non-Gaussianity}\label{sec:Non-Gaussianity-measure-Non-Gaussianity}
		The non-Gaussianity measure for the state $\ket{n,\eta}$ reads~\cite{genoni2008quantifying,albarelli2018resource,adnane2019quantum}
		\begin{equation}\label{eqn:nG-n-eta-def-formal}
			\delta_\text{nG}^{n,\eta} = \sigma_+\ln\sigma_+\ -\ \sigma_-\ln\sigma_-\ ,
		\end{equation}
		where $\sigma_\pm=\sqrt{\det\sigma}\pm\frac12$, and $\sigma\equiv\sigma^{n,\eta}$ is the covariance matrix given in \refeq{eqn:cov-mat}.

\begin{figure}
			\begin{center}
				\includegraphics[width=\linewidth]{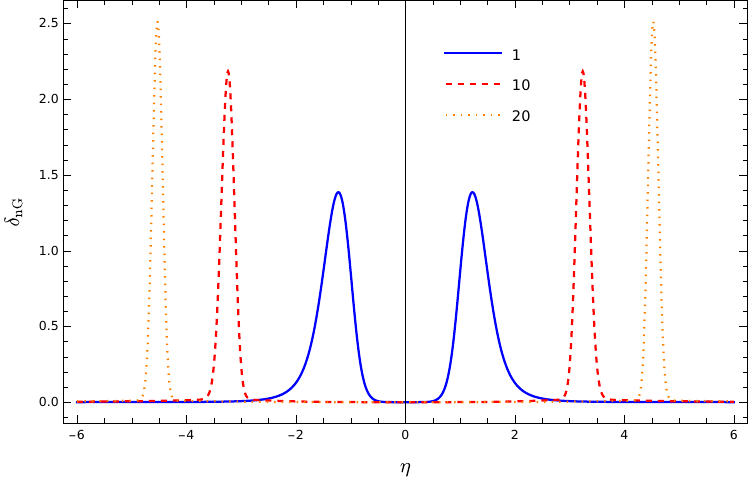}
			\end{center}
			\caption{Non-Gaussianity as a function of $\eta$ for different values of $n$. The solid (dashed, dotted) curve corresponds to $n=1$ ($n=10$, $n=20$).\label{fig:nG-eta}}
		\end{figure}

\begin{figure}
			\begin{center}
				\includegraphics[width=\linewidth]{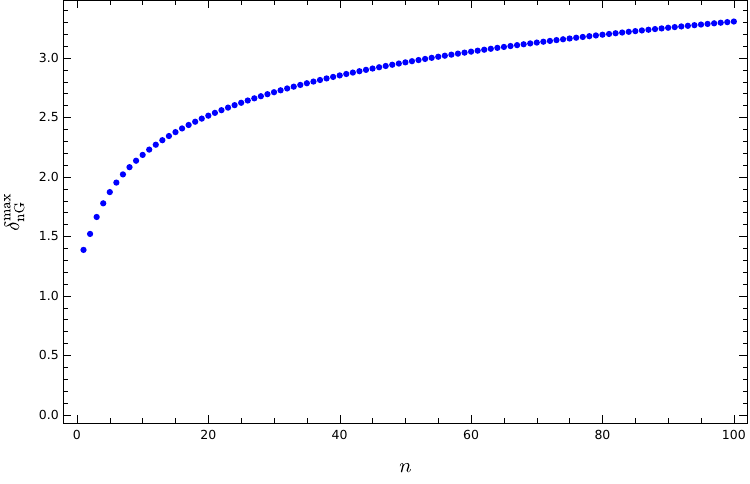}
			\end{center}
			\caption{Maximum non-Gaussianity as a function of $n$.\label{fig:nG-max-n}}
		\end{figure}

		We numerically evaluate the non-Gaussianity for different states. In \reffig{fig:nG-eta} we plot the variation of $\delta_\text{nG}^{n,\eta}$ as a function of $\eta$ for different $n$ ($n\in\{1,10,20\}$). The non-Gaussianity is an even function of $\eta$, which is obvious from the analytical expressions of the diagonal elements of the covariance matrix (\refeq{eqn:cov-mat-11-22}). Moreover, it exhibits two extrema that grow and displace away from each other as $n$ increases. The behaviour of this growth and shift as a function of $n$ can be determined. First, we find the zeros of the derivative of the function $\delta_\text{nG}^{n,\eta}$ with respect to $\eta$. This problem is relatively straightforward, as the zeros of the derivative of $\delta_\text{nG}^{n,\eta}$ are the same as those of the derivative of the determinant of the covariance matrix,
		\begin{equation}
			\frac{\partial \delta_\text{nG}^{n,\eta}}{\partial\eta} = 0 \Rightarrow
			\frac{\partial \sigma_{11}}{\partial\eta}\, \sigma_{22} + \sigma_{11}\, \frac{\partial \sigma_{22}}{\partial\eta} = 0\ .
		\end{equation}
		Solving this latter equation gives
		\begin{equation}\label{eqn:solution-diff-nG-eta=0}
			\eta_\pm = \pm\sqrt{n+\frac12}\ .
		\end{equation}
		The solutions \Refeq{eqn:solution-diff-nG-eta=0} give zero values for $\xi_{n,\eta}$,
		\begin{equation}\label{eqn:xi-max-nG}
			\xi_{n,\eta_\pm} = 0\ ,
		\end{equation}
		and the maximum non-Gaussianity expression, denoted $\delta_\text{nG}^{\text{max}}(n)$, is given by
		\begin{equation}
			\delta_{\text{nG}}^\text{max}(n) = \nu_+\ln\nu_+\ -\ \nu_-\ln\nu_-\ ,
		\end{equation}
		with $\nu_\pm$ given by
		\begin{equation}
			\nu_\pm = \sqrt{\left(n+\frac12\right)\ \frac{4n-1}{4n-2}}\ \pm\ \frac12\ .
		\end{equation}
		Clearly, the maximum non-Gaussianity is unbounded for increasing $n$, as shown in \reffig{fig:nG-max-n}.

	%====================================================
	%====================================================
	%====================================================
	\subsection{Maximally non-Gaussian states}\label{sec:max-non-Gaussianity-states}
		The state $\ket{n,\eta}$ becomes the most non-Gaussian when ${\eta=\eta_\pm=\pm\sqrt{n+\frac12}}$, as shown in \refsec{sec:Non-Gaussianity-measure-Non-Gaussianity}, namely, \refeq{eqn:solution-diff-nG-eta=0}. The state $\ket{n,\eta_\pm}\equiv\ket{\Psi_n}$ becomes dependent only on the parameter $n$, and reads:
		\begin{multline}\label{eqn:ket-n-eta-pm}
			\ket{\Psi_n}=\frac{n!}{\sqrt{2^{n-\upepsilon_n} (2n-1)!!}}\\
			\times\sum\limits_{k=0}^{\frac{n-\upepsilon_n}{2}}\frac{2^k}{\left(\frac{n-\upepsilon_n}{2}-k\right)!\ \sqrt{(2k+\upepsilon_n)!}}\ket{2k+\upepsilon_n}\ .
		\end{multline}
		The corresponding wave function simply can be written as
		\begin{equation}
			\Psi_n(x) = \frac{1}{\sqrt{\Gamma\left(n+\frac12\right)}}\ x^n\ \EXP{-\frac12 x^2}\ ,\label{eqn:Psi-n-wave-function}
		\end{equation}
		while the maximum non-Gaussian steady states of the system is
		\begin{equation}\label{eqn:maximum-non-Gaussian-steady-states}
			\hat{D}\left(\pm \frac12 s\sqrt{2n+1}\right)\ \hat{S}(s)\ket{\Psi_n}\ .
		\end{equation}

%====================================================
%====================================================
%====================================================
\section{Schr\"odinger cat states}\label{sec:cat-steady-state}
		Clearly, the states $\ket{\Psi_n}$ given in \refeq{eqn:ket-n-eta-pm} have a definite parity; they are even (odd) for even (odd) values of $n$. Moreover, the steady state does not depend on the parameter $g_1$, and it only dictates the speed at which the system approaches the steady state.

		In the following, we show that the steady state $\ket{\Psi_n}$ is a Schr\"odinger cat state with sizes solely dependent on the parameter $n$.~\footnote{The state \Refeq{eqn:ket-n-eta-pm} has previously been studied by one of the authors in~\protect\cite{brunelli2018linear}, but here we report that this state is a very good approximation to (squeezed) cat states where its size is fully controllable via the couplings $g_j$ ($j=1,\ldots,5$). This result was not studied in the aforementioned reference. Additionally, The states $\ket{\Psi_n}$ had been studied in other works, e.g.,~\protect\cite{lee2012generating, ourjoumtsev2007generation}, where the authors proposed a conditional scheme to generate them, while in our work, we propose a deterministic approach. Moreover, in this paper, we studied in more depth the non-Gaussianity of the steady states of the system and how the cat states are obtained when maximising the non-Gaussianity of the steady states, which again is a new result in this work and was not done in the previously mentioned reference.}

		We consider the squeezed Schr\"odinger cat state $\ket{\chi_{\gamma,r,\mathcal{P}}}$ defined as~\cite{schleich1991nonclassical, yurke1986generating}
		\begin{equation}\label{eqn:cat-state}
			\ket{\chi_{\gamma,r,\mathcal{P}}} = \mathcal{N}_{\gamma,r}^\mathcal{P}\Big(\hat{D}(\gamma) \hat{S}(r) \ket{0}+\mathcal{P}\hat{D}(-\gamma) \hat{S}(r) \ket{0}\Big)\ ,
		\end{equation}
		where $\ket{0}$ denotes the vacuum state, $r$ is a squeezing parameter, $\gamma$ is a real parameter quantifying the size of the cat state, and the parity parameter $\mathcal{P}$ takes two values---$\mathcal{P}=+1$ for even states and $\mathcal{P}=-1$ for odd states. Finally, $\mathcal{N}_{\gamma,r}^\mathcal{P}$ is a normalisation coefficient, which can be easily calculated as
		\begin{equation}
			\mathcal{N}_{\gamma,r}^\mathcal{P} = \frac{1}{\sqrt{2\sqrt\pi\ r\left(1+\mathcal{P}\EXP{-2\frac{\gamma^2}{r^2}}\right)}}\ .
		\end{equation}

		We shall quantify the similarity between the states $\ket{\Psi_n}$ and the Schr\"odinger cat states $\ket{\chi_{\gamma,r,\mathcal{P}}}$ by the fidelity measure~\cite{uhlmann1976transition, jozsa1994fidelity, braunstein2000criteria},
		\begin{align}
			F(n;\gamma,r,\mathcal{P}) 
			&\equiv \left|\braket{\Psi_n|\chi_{\gamma,r,\mathcal{P}}}\right|^2\ ,\nonumber\\
			&= \left|\,\int\limits_{-\infty}^\infty \Psi^*_n(x)\ \chi_{\gamma,r,\mathcal{P}}(x)\ \diff x\right|^2\ .\label{eqn:fidelity-n-gamma-r-p}
		\end{align}
		Here, $\chi_{\gamma,r,\mathcal{P}}(x)$ denotes the position-representation wave function of the state $\ket{\chi_{\gamma,r,\mathcal{P}}}$, which is given by
		\begin{equation}
			\chi_{\gamma,r,\mathcal{P}}(x) = \frac{\mathcal{N}_{\gamma,r}^\mathcal{P}}{\sqrt{r\sqrt{\pi}}}\Big(\EXP{-\frac{\left(x-\sqrt2\,\gamma\right)^2}{2r^2}}+\mathcal{P}\EXP{-\frac{\left(x+\sqrt2\,\gamma\right)^2}{2r^2}}\Big)\ .
		\end{equation}
		By evaluating the integral in the fidelity~\Refeq{eqn:fidelity-n-gamma-r-p} we arrive at the following expression,
		\begin{widetext}
			\begin{align}
				F(n;\gamma,r,\mathcal{P}) 
				&= \frac{\frac{1+(-1)^{\upepsilon_n}\mathcal{P}}{2}\ \gamma^{1-\mathcal{P}}\ 2^{n+3-\mathcal{P}}\ r^{2n+\mathcal{P}}\ \left[\Gamma\left(\frac{n+1+\frac{1-\mathcal{P}}{2}}{2}\right)\hypergeom\left(\frac{n+1+\frac{1-\mathcal{P}}{2}}{2};1-\frac{\mathcal{P}}{2};\frac{\gamma^2}{r^2(1+r^2)}\right)\right]^2}{\sqrt\pi\ \Gamma\left(n+\frac12\right)\ \left(\mathcal{P}+\EXP{2\frac{\gamma^2}{r^2}}\right) (1+r^2)^{n+1+\frac{1-\mathcal{P}}{2}}}\ .\label{eqn:fid-sqz-cat-state}
			\end{align}
		\end{widetext}

\begin{figure}
			\begin{center} 
				\includegraphics[width=\linewidth]{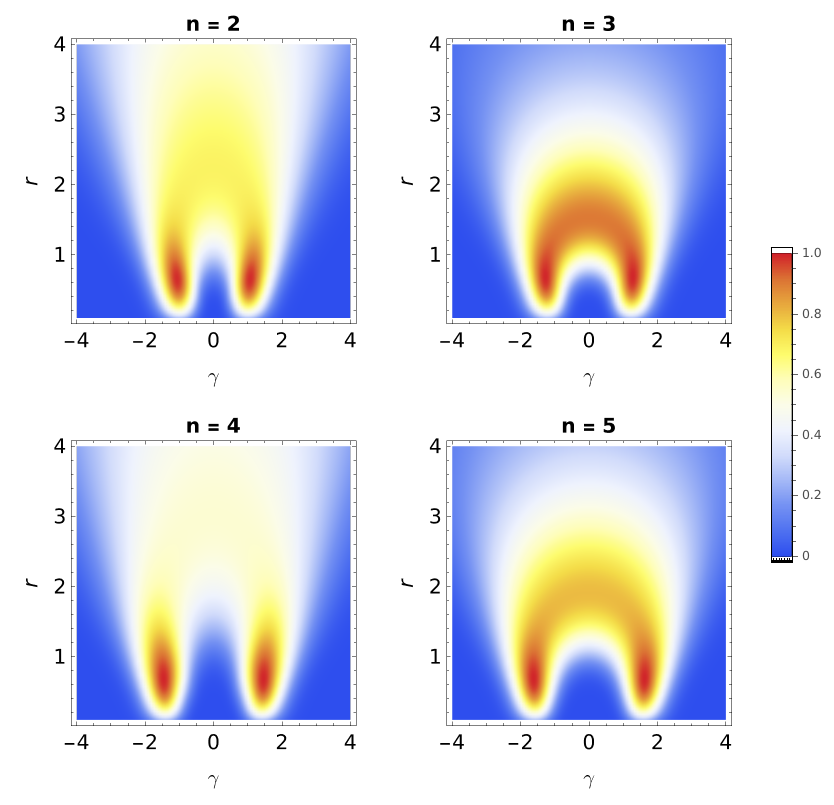}
			\end{center}
			\caption{Fidelity ${F(n;\gamma,r,\mathcal{P})}$ as a function of $\gamma$ and $r$, for different values of $n$.}
			\label{fig:fid-gamma-r}
		\end{figure}

\begin{figure}
			\begin{center} 
				\includegraphics[width=\linewidth]{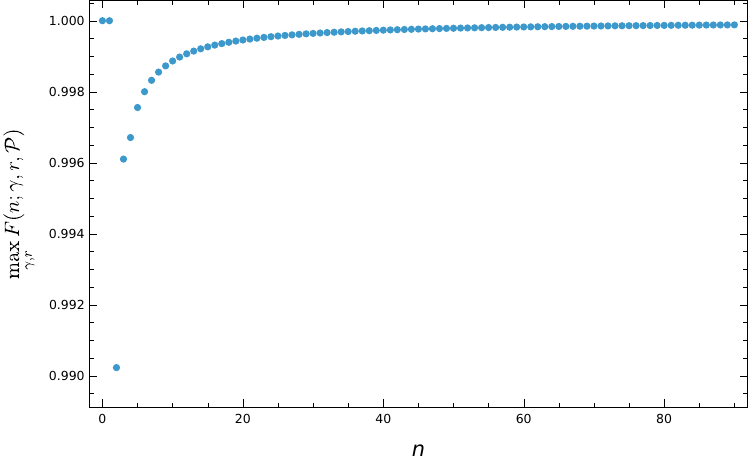}
			\end{center}
			\caption{Maximum fidelity ${\max_{\gamma,r} F(n;\gamma,r,\mathcal{P})}$ as a function of $n$, for even ($\mathcal{P}=1$) and odd ($\mathcal{P}=-1$) states.}
			\label{fig:max-fid-n-cat-states}
		\end{figure}

		Now we find the values of $\gamma$ and $r$ that lead to a maximal fidelity $F(n;\gamma,r,\mathcal{P})$ for a fixed $n$. \reffig{fig:fid-gamma-r} shows the variation of the fidelity $F(n;\gamma,r,\mathcal{P})$ as a function of the cat size, $\gamma$, and the squeezing parameter, $r$. The fidelity reaches nearly unity for particular values of $\gamma$ and $r$. In \reffig{fig:max-fid-n-cat-states} we plot the variation of this maximum, as a function of $n$. We see that ${\max_{\gamma,r} F(n;\gamma,r,\mathcal{P})}$ approximately reaches unit value for increasing $n$. Even for relatively small values of $n$ ($n\ge 3$), the fidelity is above $0.996$. For each value of $n$, the maximum fidelity corresponds to a pair of optimal values for the squeezing $r$ and cat size $\gamma$. Moreover, the parameter domain with high fidelity is located at approximately similar values of $r$ for different values of $n$.

\begin{figure}
			\begin{center} 
				\includegraphics[width=\linewidth]{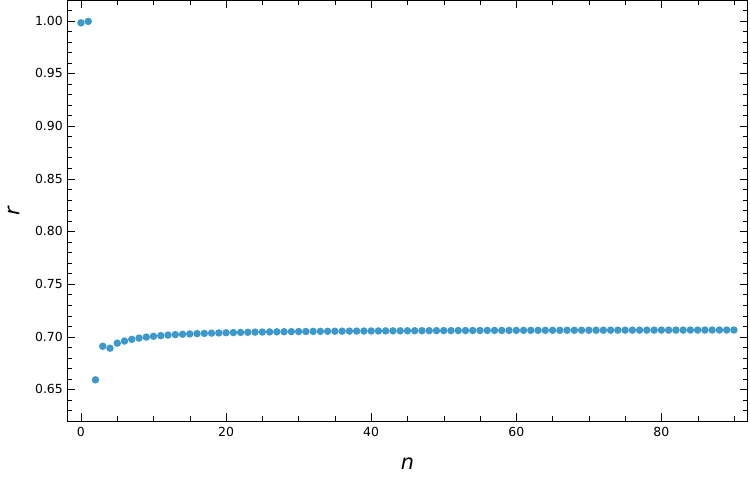}
			\end{center}
			\caption{Variation of the squeezing $r$ at maximum fidelity as a function of $n$.}
			\label{fig:max-fid-n-cat-states-r}
		\end{figure}

\begin{figure}
			\begin{center} 
				\includegraphics[width=\linewidth]{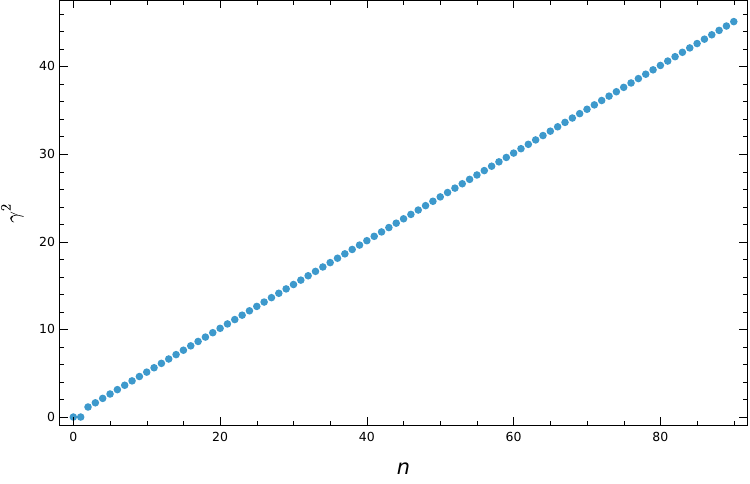}
			\end{center}
			\caption{Variation of the cat size $\gamma^2$ at maximum fidelity as a function of $n$.}
			\label{fig:max-fid-n-cat-states-gamma}
		\end{figure}

		In Figs.~\ref{fig:max-fid-n-cat-states-r} and~\ref{fig:max-fid-n-cat-states-gamma} we respectively plot the optimal values of $r$ and $\gamma$ that correspond to each point in the plot of ${\max_{\gamma,r} F(n;\gamma,r,\mathcal{P})}$ of \reffig{fig:max-fid-n-cat-states}. While the squeezing at every value of $n$ remains around $r=\frac{1}{\sqrt2}$ or equivalently $3~\dB$ of $x$-squeezing---except for $n=0$ or $n=1$, we notice that the cat's size increases with $n$. Namely, the square of the cat size is linear in $n$ with equation
		\begin{equation}
			\gamma^2 \approx \frac12\ n\ +\ 0.627\ .
		\end{equation}

		In conclusion, it is possible to obtain the squeezed Schr\"odinger cat states \Refeq{eqn:cat-state}, with very high fidelities exceeding $\num{0.996}$ for $n\ge 3$. The size of the cat state is determined by the parameter $n$, which in turn is fixed by choosing the right ratios of the linear and quadratic couplings, as already shown in \refsec{sec:displaced-squeezed-superposition}.

%====================================================
%====================================================
%====================================================
\section{Wigner function and negativity}\label{sec:wigner-function-and-negativity}
	%====================================================
	%====================================================
	%====================================================
	\subsection{Analytical expression of Wigner function}\label{sec:wigner}
		Here, we calculate the Wigner function that corresponds to the steady state of the system governed by the master equation~\Refeq{eqn:master-equation}, i.e. the state with wave function $\varphi_{n,s,x_0}(x)$.
		
		We denote by $\mathcal{W}_{n,s,x_0}$ and $W_{n,\eta}$ the Wigner functions of $\ket{\varphi_{n,s,x_0}}$ and $\ket{n,\eta}$ respectively. Since these ket vectors are related by
		\begin{equation}
			\ket{\varphi_{n,s,x_0}}=\hat{D}\left(\frac{x_0}{\sqrt2}\right)\hat{S}(s)\ket{n,\eta}\ ,
		\end{equation}
		then the corresponding Wigner functions satisfy the equation~\cite{albarelli2018resource}
		\begin{equation}
			\mathcal{W}_{n,s,x_0}(x,p)=W_{n,\eta}(\tilde{x},\tilde{p})\ ,
		\end{equation}
		where the new variables $\tilde{x}$ and $\tilde{p}$ are deduced from the transformation of the operators $\hat{x}$ and $\hat{p}$ by the Gaussian operator $\hat{D}\left(\frac{x_0}{\sqrt2}\right)\hat{S}(s)$ as shown in the following. We have
		\begin{subequations}
			\begin{align}
				\hat{D}\left(\frac{x_0}{\sqrt2}\right)\hat{S}(s)\ \hat{x}\ S^\dagger(s)\hat{D}\left(\frac{x_0}{\sqrt2}\right)^\dagger&=\frac{1}{s}\ \hat{x}\ -\ \eta\ ,\\
				\hat{D}\left(\frac{x_0}{\sqrt2}\right)\hat{S}(s)\ \hat{p}\ S^\dagger(s)\hat{D}\left(\frac{x_0}{\sqrt2}\right)^\dagger&=s\ \hat{p}\ .
			\end{align}
		\end{subequations}
		Consequently, the expressions of the new variables $\tilde{x}$ and $\tilde{p}$ are written in a similar form as
		\begin{subequations}
			\begin{align}
				\tilde{x}	&=	\frac{1}{s}\ x\ -\ \eta\ ,\\
				\tilde{p}	&=	s\ p\ ,
			\end{align}
		\end{subequations}
		and the Wigner function is given by
		\begin{equation}\label{eqn:Wigner-phi-n-s-x0}
			\mathcal{W}_{n,s,x_0}(x,p)=W_{n,\eta}\left(\frac{1}{s}\ x\ -\ \eta,s\ p\right)\ .
		\end{equation}
		
		In the following we find the Wigner function $W_{n,\eta}$. It is defined by the expression~\cite{schleich2001quantum,wigner1932quantum}
		\begin{equation}
			W_{n,\eta}(x,p)=\frac{1}{2\pi}\ \int_{-\infty}^\infty\psi_{n,\eta}\left(x-\frac{y}{2}\right)\psi^*_{n,\eta}\left(x+\frac{y}{2}\right)\EXP{ipy}\ \diff y\ .
		\end{equation}
		By using the wave function expression given in \refeq{eqn:wave-function-psi}, making the change of variables ${y\rightarrow 2y}$, and after a few simplifications, the Wigner function is now written as
		\begin{multline}
			W_{n,\eta}(x,p)=\frac{(-1)^n}{\pi\Gamma\left(n+\frac12\right)\hypergeom\left(-n;\frac12;-\xi_{n,\eta}^2\right)}\\
			\times \EXP{-x^2-p^2} \int_{-\infty}^\infty\Big(y^2-(x+\xi_{n,\eta})^2\Big)^n\EXP{-(y-ip)^2}\diff y\ .
		\end{multline}
		The integral can be calculated by, first,  using the binomial expansion,
		\begin{equation}
			\Big(y^2-(x+\xi_{n,\eta})^2\Big)^n=\sum\limits_{k=0}^n\binom{n}{k} (-1)^{n-k} (x+\xi_{n,\eta})^{2(n-k)} y^{2k}\ ,
		\end{equation}
		then by evaluating the integral
		\begin{equation}
			\int\limits_{-\infty}^\infty y^{2k}\EXP{-(y-ip)^2}\diff y=\Gamma\left(k+\frac12\right) \hypergeom\left(-k;\frac12;p^2\right)\ ,
		\end{equation}
		one can arrive at the following expression for the Wigner function for the state $\ket{n,\eta}$
		\begin{multline}
			W_{n,\eta}(x,p)=\frac{\EXP{-x^2-p^2}}{\pi\Gamma\left(n+\frac12\right)\hypergeom\left(-n;\frac12;-\xi_{n,\eta}^2\right)}\\
			\times\sum\limits_{k=0}^n\binom{n}{k} (-1)^{k} \Gamma\left(k+\frac12\right)\\
			\times (x+\xi_{n,\eta})^{2(n-k)} \hypergeom\left(-k;\frac12;p^2\right)\ .\label{eqn:wigner-n-eta}
		\end{multline}
		Finally, taking into account \refeq{eqn:Wigner-phi-n-s-x0}, one obtains the Wigner function for the state $\ket{\varphi_{n,s,x_0}}$,
		\begin{multline}
			\mathcal{W}_{n,s,x_0}(x,p)=\frac{\EXP{\frac{-1}{s^2}\left(x-x_0\right)^2-s^2p^2}}{\pi\Gamma\left(n+\frac12\right)\hypergeom\left(-n;\frac12;-\xi_{n,\eta}^2\right)}\\
			\times \sum\limits_{k=0}^n\binom{n}{k} (-1)^{k} \Gamma\left(k+\frac12\right)\\
			\times \left(\frac{1}{s} x-\frac{n+\frac12}{\eta}\right)^{2(n-k)} \hypergeom\left(-k;\frac12;s^2p^2\right)\ .\label{eqn:wigner-n-s-x0}
		\end{multline}

\begin{figure}[htb]
			\begin{center}
				\includegraphics[width=\linewidth]{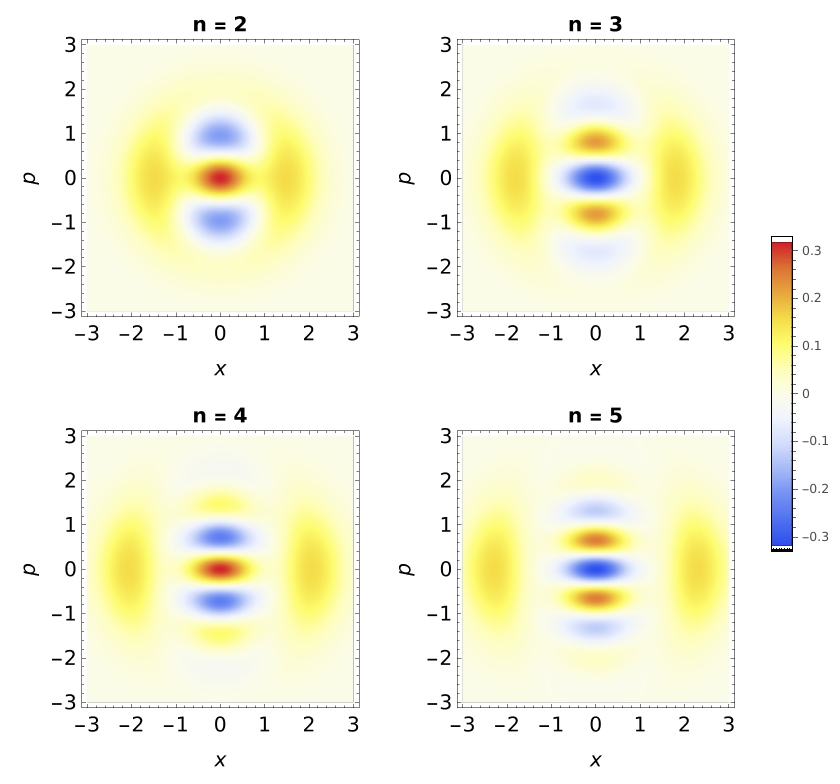}
				\caption{The Wigner function $W_{n,\eta}$ in phase space. The value of $\eta$ for each plot is $\eta=\sqrt{n+\frac12}$.}\label{fig:wigner-n-eta-plots}
			\end{center}
		\end{figure}

		In the following, we consider the Wigner function given in \refeq{eqn:wigner-n-eta} and show its plot for a set of values for $n$ and $\eta$. The plots for the Wigner function $\mathcal{W}_{n,s,x_0}$ are a displaced and squeezed form of the plots for $W_{n,\eta}$, while the non-classical features will be the same for both. \reffig{fig:wigner-n-eta-plots} shows the variations of the Wigner $W_{n,\eta}$ in phase space. It is clear that these states are highly non-classical, and that their Wigner functions are non-Gaussian.

	%===========================================
	%===========================================
	\subsection{Wigner logarithmic negativity}\label{sec:mana}
		We study the negativity of the Wigner function for the steady state of the system. Since the negativity of the Wigner function for a quantum state is not affected by the application of Gaussian operations on the state~\cite{albarelli2018resource}, it is sufficient to study the negativity of the Wigner function for the state $\ket{n,\eta}$, i.e. \refeq{eqn:wigner-n-eta}.

\begin{figure}[hb]
			\begin{center}
				\includegraphics[width=\linewidth]{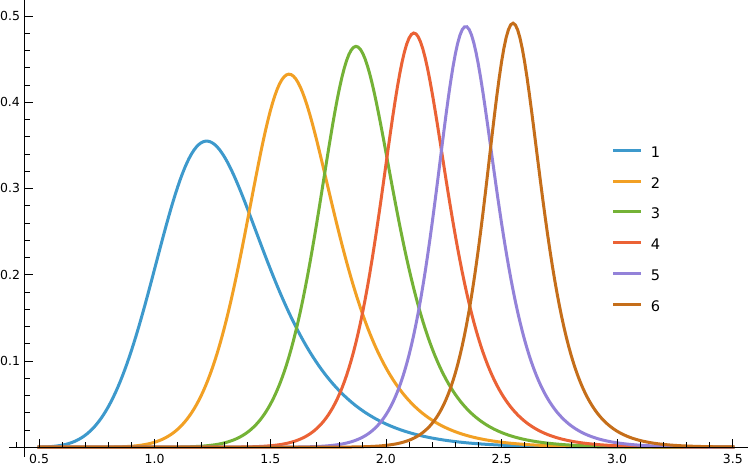}
				\caption{WLN as a function of the ratio $\eta$ for different values of $n$ ($n=1,\ldots,6$).}\label{fig:mana-eta}
			\end{center}
		\end{figure}

\begin{figure}[htb]
			\begin{center}
				\includegraphics[width=\linewidth]{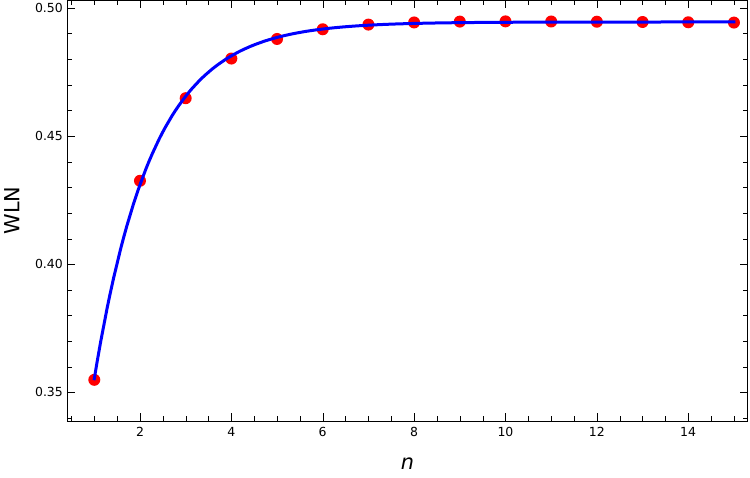}
				\caption{$\max\limits_{\eta>0}\mathcal{M}(n,\eta)$ as a function of $n$. The dots represent the maximum values of the WLN, and the numbers inside the bars are the corresponding values of $\eta$ where the WLN is at its maximum. The blue curve is the exponential fit function given in \refeq{eqn:exp-fit-function}.}\label{fig:mana-eta-max}
			\end{center}
		\end{figure}

		In the following, we will study properties of the Wigner logarithmic negativity (WLN)~\cite{albarelli2018resource} and the negative extremum of the Wigner function. The WLN associated to the Wigner function $W_{n,\eta}$ is defined by~\cite{albarelli2018resource}
		\begin{equation}
			\mathcal{M}(n,\eta)=\log\iint\limits_{\mathds{R}^2}\left|W_{n,\eta}(x,p)\right|\diff x\diff p\ .\label{eqn:mana-n-eta}
		\end{equation}
		In \reffig{fig:mana-eta} we plot the variations of the WLN $\mathcal{M}$ as a function of the parameter $\eta$ for different values of $n$. We notice two behaviours of the WLN: (i)~the function $\mathcal{M}$ has a maximum at one particular value of $\eta$, and (ii)~this extremum saturates with increasing values of $n$. Following this observation, we plot, in \reffig{fig:mana-eta-max}, the variations of the maximum of the WLN, $\max\limits_{\eta>0}\mathcal{M}(n,\eta)$, as a function of $n$. We tried an exponential fitting of the results and found
		\begin{equation}
			\max\limits_{\eta>0}\mathcal{M}(n,\eta) \approx 0.4945-0.3051\,\EXP{-0.7848n}\ .\label{eqn:exp-fit-function}
		\end{equation}
		It is clear that $\ket{n,\eta}$ becomes a maximally non-classical state starting from relatively small values of $n$ ($n\ge 6$). Moreover, we noticed that the values of $\eta$ where the WLN is maximal---see \reffig{fig:mana-eta-max}---write in terms of $n$ as
		\begin{equation}
			\eta = \sqrt{n+\frac12}\ .
		\end{equation}
		This is in agreement with the result of \refsec{sec:Non-Gaussianity-measure}, where we studied the maximally non-Gaussian steady states.

% ====================================
% ========Conclusions and outlook=========
% ====================================
\section{Conclusions and outlook}\label{sec:conclusion}
	In this paper, we have introduced a scheme for the deterministic preparation of a family of non-Gaussian states, including squeezed Schr\"odinger cat states, using reservoir engineering technique. The scheme relies on two linearly and quadratically coupled bosonic modes, one of which is dissipatively coupled to a cold reservoir. After adiabatic elimination of this auxiliary mode, the remaining oscillator evolves under a purely dissipative master equation.  We have established the conditions on the coupling parameters under which the dynamics admits a pure steady state and obtained an analytical solution of the master equation. We have shown that this steady state takes the form of a displaced and squeezed finite superposition of Fock states. Specifically, the finite superposition part of the steady state---$\ket{n,\eta}$, given in \refeq{eqn:ket-n-eta}---contains Fock states up to $\ket{n}$ and depends only on the integer $n$ and the ratio $\eta=x_0/s$. The steady state of the studied system is obtained in a deterministic fashion. Numerical simulations confirm that it is reached with unit fidelity even when the system is initialised in a thermal state. Furthermore, we derived analytical expressions for the non-Gaussianity and the Wigner function of the steady state. Our results reveal that the maximally non-Gaussian steady states provide excellent approximations to squeezed Schrödinger cat states and simultaneously exhibit maximal Wigner negativity. The fidelities exceed $0.996$ already for small values of $n$ ($n\ge3$) and continue increasing as $n$ increases. The optimal squeezing stays around $r\approx1/\sqrt{2}$ ($3~\dB$), while the square of the cat size grows linearly with $n$.

	A key advantage of our scheme is its deterministic and unconditional nature. Since the target state is the unique steady state of the dynamics, the system converges towards it independently of its initial conditions, as confirmed numerically for a thermal initial state. No measurements, feedback, or post-selection are required, thereby avoiding the trade-off between preparation fidelity and success probability inherent to probabilistic protocols discussed in the introduction. The size of the generated cat states is solely controlled through the linear and quadratic couplings, which are usually experimentally tunable. Moreover, since Gaussian operations preserve the Wigner negativity, the generated states can be further displaced, squeezed or processed without losing their non-Gaussian character. In fact, any other Gaussian operation can be embedded in the steady state of the system, by choosing the right coupling ratios. This flexibility was illustrated in \refsec{sec:system-dynamics} where we demonstrated that previously applied squeezing and displacement transformations can be counteracted by appropriately modifying the coupling parameters.

	An additional advantage of the proposed scheme is that it relies only on an engineered dissipator comprising linear and quadratic terms in the ladder operators of a bosonic mode, making it independent of any specific physical platform. Such dissipators can be engineered in a variety of physical systems, including, but not limited to, optomechanical, superconducting and microwave systems. A natural extension of this work is to propose and thoroughly study a concrete physical implementation of the scheme and assess its robustness against realistic noise sources and imperfections. More broadly, our work provides a route towards the deterministic generation of non-Gaussian quantum states, including cat states, which are key resources for quantum technologies such as quantum computing, quantum communication and teleportation, and quantum metrology and sensing.

% ====================================
% ========Acknowledgements=============
% ====================================

\section*{Acknowledgements}
	O.H. acknowledges support from the Algerian Directorate-General for Scientific Research and Technological Development (DGRSDT) (Project Code PRFU/B00L02UN260120230003).

%====================================================
%====================================================
%====================================================
	%----------
	%references
	%----------
	\bibliography{references}
\end{document}